\documentstyle[12pt]{article}
\def\emline#1#2#3#4#5#6{%
       \put(#1,#2){\special{em:moveto}}%
       \put(#4,#5){\special{em:lineto}}}
\def\newpic#1{}

\catcode`\@=11
\def\marginnote#1{}

\newcount\hour
\newcount\minute
\newtoks\amorpm
\hour=\time\divide\hour by60
\minute=\time{\multiply\hour by60 \global\advance\minute by-\hour}
\edef\standardtime{{\ifnum\hour<12 \global\amorpm={am}%
        \else\global\amorpm={pm}\advance\hour by-12 \fi
        \ifnum\hour=0 \hour=12 \fi
        \number\hour:\ifnum\minute<10 0\fi\number\minute\the\amorpm}}
\edef\militarytime{\number\hour:\ifnum\minute<10 0\fi\number\minute}

\def\draftlabel#1{{\@bsphack\if@filesw {\let\thepage\relax
   \xdef\@gtempa{\write\@auxout{\string
      \newlabel{#1}{{\@currentlabel}{\thepage}}}}}\@gtempa
   \if@nobreak \ifvmode\nobreak\fi\fi\fi\@esphack}
        \gdef\@eqnlabel{#1}}
\def\@eqnlabel{}
\def\@vacuum{}
\def\draftmarginnote#1{\marginpar{\raggedright\scriptsize\tt#1}}

\def\draft{\oddsidemargin -.5truein
        \def\@oddfoot{\sl preliminary draft \hfil
        \rm\thepage\hfil\sl\today\quad\militarytime}
        \let\@evenfoot\@oddfoot \overfullrule 3pt
        \let\label=\draftlabel
        \let\marginnote=\draftmarginnote
   \def\@eqnnum{{\rm (\theequation)}\rlap{\kern\marginparsep\tt\@eqnlabel}%
\global\let\@eqnlabel\@vacuum}  }

\def\bea{\begin{eqnarray}}
\def\eea{\end{eqnarray}}

\def\beq{\begin{equation}}
\def\eeq{\end{equation}}
\def\ba{\beq\new\begin{array}{c}}
\def\ea{\end{array}\eeq}
\def\be{\ba}
\def\ee{\ea}

\def\Tr{{\rm Tr}}

\makeatletter
\newdimen\normalarrayskip              
\newdimen\minarrayskip                 
\normalarrayskip\baselineskip
\minarrayskip\jot
\newif\ifold             \oldtrue            \def\new{\oldfalse}
\def\arraymode{\ifold\relax\else\displaystyle\fi} 
\def\eqnumphantom{\phantom{(\theequation)}}     
\def\@arrayskip{\ifold\baselineskip\z@\lineskip\z@
     \else
     \baselineskip\minarrayskip\lineskip2\minarrayskip\fi}
\def\@arrayclassz{\ifcase \@lastchclass \@acolampacol \or
\@ampacol \or \or \or \@addamp \or
   \@acolampacol \or \@firstampfalse \@acol \fi
\edef\@preamble{\@preamble
  \ifcase \@chnum
     \hfil$\relax\arraymode\@sharp$\hfil
     \or $\relax\arraymode\@sharp$\hfil
     \or \hfil$\relax\arraymode\@sharp$\fi}}
\def\@array[#1]#2{\setbox\@arstrutbox=\hbox{\vrule
     height\arraystretch \ht\strutbox
     depth\arraystretch \dp\strutbox
     width\z@}\@mkpream{#2}\edef\@preamble{\halign
\noexpand\@halignto
\bgroup \tabskip\z@ \@arstrut \@preamble \tabskip\z@ \cr}%
\let\@startpbox\@@startpbox \let\@endpbox\@@endpbox
  \if #1t\vtop \else \if#1b\vbox \else \vcenter \fi\fi
  \bgroup \let\par\relax
  \let\@sharp##\let\protect\relax
  \@arrayskip\@preamble}
\def\eqnarray{\stepcounter{equation}%
              \let\@currentlabel=\theequation
              \global\@eqnswtrue
              \global\@eqcnt\z@
              \tabskip\@centering
              \let\\=\@eqncr
              $$%
 \halign to \displaywidth\bgroup
    \eqnumphantom\@eqnsel\hskip\@centering
    $\displaystyle \tabskip\z@ {##}$%
    \global\@eqcnt\@ne \hskip 2\arraycolsep
         $\displaystyle\arraymode{##}$\hfil
    \global\@eqcnt\tw@ \hskip 2\arraycolsep
         $\displaystyle\tabskip\z@{##}$\hfil
         \tabskip\@centering
    &{##}\tabskip\z@\cr}
\begingroup\ifx\undefined\newsymbol \else\def\input#1 {\endgroup}\fi
\input amssym.def \relax
\input amssym
\newfont{\hr}{msbm10}
\newfont{\ams}{msam10}

\font\numbers=cmss12
\font\upright=cmu10 scaled\magstep1
\def\stroke{\vrule height8pt width0.4pt depth-0.1pt}
\def\topfleck{\vrule height8pt width0.5pt depth-5.9pt}
\def\botfleck{\vrule height2pt width0.5pt depth0.1pt}
\def\Zmath{\vcenter{\hbox{\numbers\rlap{\rlap{Z}\kern 0.8pt\topfleck}\kern
2.2pt
                   \rlap Z\kern 6pt\botfleck\kern 1pt}}}
\def\Qmath{\vcenter{\hbox{\upright\rlap{\rlap{Q}\kern
                   3.8pt\stroke}\phantom{Q}}}}
\def\Nmath{\vcenter{\hbox{\upright\rlap{I}\kern 1.7pt N}}}
\def\Cmath{\vcenter{\hbox{\upright\rlap{\rlap{C}\kern
                   3.8pt\stroke}\phantom{C}}}}
\def\Rmath{\vcenter{\hbox{\upright\rlap{I}\kern 1.7pt R}}}
\def\Z{\ifmmode\Zmath\else$\Zmath$\fi}
\def\Q{\ifmmode\Qmath\else$\Qmath$\fi}
\def\N{\ifmmode\Nmath\else$\Nmath$\fi}
\def\C{\ifmmode\Cmath\else$\Cmath$\fi}
\def\R{\ifmmode\Rmath\else$\Rmath$\fi}

\newcounter{app}

\def\app{\setcounter{equation}{0}
\def\theequation{\Alph{app}.\arabic{equation}}\par
   \addvspace{4ex}
   \@afterindentfalse
  \secdef\@app\@dapp}

\newcommand\@app{\@startsection {app}{1}{0ex}%
                                   {-3.5ex \@plus -1ex \@minus -.2ex}%
                                   {2.3ex \@plus.2ex}%
                                   {\normalfont\Large\bf}}
\def\@dapp#1{%
{\parindent \z@ \raggedright  \bf #1}\par\nobreak}
\def\l@app#1#2{\ifnum \c@tocdepth >\z@
    \addpenalty\@secpenalty
    \addvspace{1.0em \@plus\p@}%
    \setlength\@tempdima{8em}%
    \begingroup
      \parindent \z@ \rightskip \@pnumwidth
      \parfillskip -\@pnumwidth
      \leavevmode \bfseries
      \advance\leftskip\@tempdima
      \hskip -\leftskip
      #1\nobreak\hfil \nobreak\hb@xt@\@pnumwidth{\hss #2}\par
    \endgroup\fi}
\newcounter{sapp}[app]

\def\sapp{\def\theequation{\Alph{app}.\arabic{equation}}
\par
\@afterindentfalse
  \secdef\@sapp\@dsapp}
\newcommand{\@sapp}{\@startsection{sapp}{2}{\z@}%
                                     {-3.25ex\@plus -1ex \@minus -.2ex}%
                                     {1.5ex \@plus .2ex}%
                                     {\normalfont\large\bfseries}}

\def\@dsapp#1{%
{\parindent \z@ \raggedright  \bf #1
}\par\nobreak}
\newcommand{\l@sapp}{\@dottedtocline{2}{1.5em}{2.3em}}

\def\Tr{{\rm Tr}}

\def\2{{1\over 2}}
\def\N2{${\cal N}=2$}

\def\be{ \begin{eqnarray} }
\def\ee{ \end{eqnarray} }

\def\bea{\begin{eqnarray}}
\def\eea{\end{eqnarray}}

\def\beq{\begin{equation}}
\def\eeq{\end{equation}}
\def\ba{\beq\new\begin{array}{c}}
\def\ea{\end{array}\eeq}
\def\be{\ba}
\def\ee{\ea}

\begin{document}
\begin{flushright}
ITEP/TH-78/98\\
FIAN/TD-09/98\\
hepth/9902030
\end{flushright}
\vspace{0.5cm}
\begin{center}
{\LARGE \bf Solutions to the reflection equation\\ and integrable systems\\}
\vspace{0.2cm}
{\LARGE \bf for N=2 SQCD with classical groups}
\vspace{0.5cm}

\setcounter{footnote}{1}
\def\thefootnote{\fnsymbol{footnote}}
{\Large A.Gorsky\footnote{ITEP,
Moscow 117259, Russia; e-mail address:
gorsky@vx.itep.ru}, A.Mironov\footnote{Theory
Department, Lebedev Physics Institute, Moscow
~117924, Russia; e-mail address: mironov@lpi.ru}\footnote{ITEP,
Moscow 117259, Russia; e-mail address:
mironov@itep.ru}}\\
\end{center}
\bigskip
\begin{quotation}
\noindent
Integrable systems underlying the Seiberg-Witten solutions for
the $N=2$ SQCD with gauge groups $SO(n)$ and $Sp(n)$ are
proposed. They are described by the inhomogeneous XXX spin chain
with specific boundary conditions given by reflection
matrices. We attribute reflection matrices to orientifold planes in
the brane construction and briefly discuss its possible deformations.
\end{quotation}
\setcounter{footnote}{0}
\paragraph{{\large\bf Introductory remarks.}}
Since N.Seiberg and E.Witten proposed their anzatz \cite{SW} for solving
$N=2$ SUSY gauge theories, there have been a lot of attempts to realize the
structures behind it, in order to get any kind of understanding
and, after all, derivation. In particular, one of the important structures
that underlines the Seiberg-Witten (SW) anzatz and reflects its symmetry
properties is integrability \cite{GKMMM}. Concretely, in \cite{GKMMM} it has
been shown that the SW solution of the pure gauge N=2 SUSY theories with
$SU(N_c)$ gauge group can be described in the framework of the periodic Toda
chain with $N_c$ sites.

Since then, there have been a lot of different examples of the correspondence
(SW solution $\longleftrightarrow$ integrable system) considered
\cite{MW,GMMM,int,Cal,hd,GGM1,Mir97}. The list of examples includes $4d$,
$5d$ and $6d$ theories with matter hypermultiplets in adjoint or fundamental
representations included. However, all the examples from this extensive list
mainly dealt with the $SU(N_c)$ group. Not much has been known of other
groups up to the recent time. In fact, the first paper dealt with integrable
structures for other classical groups was \cite{MW}. The authors of \cite{MW}
considered the pure gauge theory with gauge group $G$ that is one of the
classical groups $SO(n)$ or $Sp(n)$ and demonstrated that the
corresponding SW anzatz can be described by the Toda chain associated with
the root system of the dual affine algebra ${\widehat {\cal G}}^{\vee}$ (one
should also specifically match the rank of this algebra, see below).

This result has been recently generalized to the theories with adjoint
matter, which are described by the elliptic Calogero-Moser model \cite{Cal}.
For these systems, Lax representation with the spectral parameter has been
constructed for the classical groups other than $SU(N_c)$ in \cite{HP} (and
the proper brane picture has been suggested in \cite{Ur,Yo}). However,
including the fundamental matter for all thee classical groups has remained a
problem. Indeed, the theories of such a type are described by spin chains
that generalize the different (i.e. of non-Calogero type), $2\times 2$ Lax
representation of the Toda chain \cite{GMMM,GGM1}. This representation
is related to the root system not immediately being rather a building block
of the final answer associated with some Dynkin diagram. The whole
construction also requires some specific boundary conditions typically
given by reflection matrices that solve the so called reflection equation
\cite{Skl88}. On the other hand, introducing the boundary conditions and
the corresponding reflection matrices is the only new thing as compared with
the $SU(N_c)$ case. Demonstration of this fact is the main purpose of
the present paper. We will show that the SW solutions for theories with
fundamental matter are {\it always} (for any classical group) governed by
the inhomogeneous XXX chain so that the group is encoded completely in the
boundary conditions (that can be read off immediately from the corresponding
Dynkin diagram).

There is an alternative way to deal with SUSY gauge theories --
that is, to use the brane pictures \cite{W}. Such pictures for the theories
with all classical groups and fundamental matter have been known for a while.
In particular, for description of the theories with orthogonal and symplectic
gauge groups, one needs, besides branes to introduce an orientifold
\cite{orien}. However, so far there has not been presented any integrable
system for the theories with orientifold. As we show in the paper, this is
just the reflection matrix to be introduced into the spin chain,
that is equivalent to adding the orientifold to the brane system.
We discuss the correspondence of the spin chain and the brane picture at the
end of the paper.

\paragraph{{\large\bf Pure gauge $SU(N_c)$ theory.}}
We begin with discussing the standard construction of the SW anzatz for the
pure gauge $SU(N_c)$ theory within the integrable
framework \cite{GKMMM}. The most important result
of \cite{SW}, from this point of view, is that the moduli space of
vacua and low energy effective action in SYM theories
are completely given by the following input data:
\begin{itemize}
\item
Riemann surface ${\cal C}$
\item
moduli space ${\cal M}$ (of the curves ${\cal C}$)
\item
meromorphic 1-form $dS$ on ${\cal C}$
\end{itemize}

Now we describe how this data can be obtained from an integrable system.
We start with the periodic Toda chain which describes the pure gauge
$SU(N_c)$ theory.
The periodic Toda chain of length $N_c$ is given by the
$2\times 2$ Lax matrices
\be\label{TodaLax}
L_i=\left(
\begin{array}{cc}
\lambda+p_i&e^{q_i}\\
-e^{-q_i}&0
\end{array}
\right)
\ee
The linear problem in the Toda chain has the following form
\be\label{lproblem}
L_i(\lambda)\Psi_i(\lambda)=\Psi_{i+1}(\lambda)
\ee
where $\Psi_i(\lambda)$ is the two-component Baker-Akhiezer function.

One also needs to consider proper
boundary conditions. These are periodic boundary conditions in the $SU(N_c)$
case. The periodic boundary conditions are easily formulated in terms
of the Baker-Akhiezer function and read as
\be\label{pbc}
\Psi_{i+N_c}(\lambda)=w\Psi_{i}(\lambda)
\ee
where $w$ is a free parameter (diagonal matrix).
The Toda chain with these boundary conditions can be naturally associated with
the Dynkin diagram of the group $A_{N_c-1}^{(1)}$, see Fig.1.


\begin{figure}[t]
\special{em:linewidth 0.4pt}
\unitlength 1.00mm
\linethickness{0.4pt}
\begin{picture}(99.00,93.05)
\put(16.33,80.00){\circle{2.00}}
\emline{17.33}{80.00}{1}{28.00}{80.00}{2}
\put(29.00,80.00){\circle{2.00}}
\put(43.00,80.00){\circle{2.00}}
\put(54.00,80.00){\circle{0.67}}
\put(58.00,80.00){\circle{0.67}}
\put(62.67,80.00){\circle{0.67}}
\put(73.67,80.00){\circle{2.00}}
\put(85.33,80.00){\circle{2.00}}
\put(98.00,80.00){\circle{2.00}}
\emline{30.33}{80.00}{3}{42.00}{80.00}{4}
\emline{75.00}{80.00}{5}{84.33}{80.00}{6}
\emline{86.33}{80.00}{7}{97.33}{80.00}{8}
\put(58.00,92.00){\circle{2.11}}
\put(58.33,66.33){\makebox(0,0)[cc]{Fig.1 Dynkin diagram for $A_n^{(1)}$}}
\emline{59.00}{92.00}{9}{97.67}{80.67}{10}
\emline{57.00}{92.00}{11}{17.00}{80.67}{12}
\end{picture}
\vspace{-6cm}
\end{figure}

One can introduce the transfer matrix shifting $i$ to $i+N_c$
\be\label{Tmat}
T(\lambda)\equiv L_{N_c}(\lambda)\ldots L_1(\lambda)
\ee
Now the periodic boundary conditions are encapsulated in
the spectral curve equation
\be\label{specurv}
\det (T(\lambda)-w\cdot {\bf 1})=0
\ee
The transfer matrix generates a complete set of integrals of motion.
Note that the Riemann surface ${\cal C}$ of the SW data is nothing but
the spectral curve.

Integrability of the Toda chain follows from
{\it quadratic} r-matrix relations (see, e.g. \cite{FT})
\be\label{quadr-r}
\left\{L_i(\lambda)\stackrel{\otimes}{,}L_j(\lambda')\right\} =
\delta_{ij}
\left[ r(\lambda-\lambda'),\ L_i(\lambda)\otimes L_i(\lambda')\right]
\ee
with the rational $r$-matrix
\be\label{rat-r}
r(\lambda) = \frac{1}{\lambda}\sum_{a=1}^3 \sigma^a\otimes \sigma^a
\ee
where $\sigma^a$ are the standard Pauli matrices.

The crucial property of this relation is that it
is multiplicative and any product like (\ref{Tmat})
satisfies the same relation
\be\label{Tbr}
\left\{T(\lambda)\stackrel{\otimes}{,}T(\lambda')\right\} =
\left[ r(\lambda-\lambda'),\
T(\lambda)\otimes T(\lambda')\right]
\ee
and, therefore, traces of the transfer matrices given at different
$\lambda$ are commuting giving rise to a series of conservation laws
(= Hamiltonians).

The spectral curve (\ref{specurv}) can be presented
in more explicit terms:
\be\label{Todasc}
w^2-\Tr T(\lambda) w+1=0
\ee
since $\det T(\lambda)=1$.
Transfer matrix $\Tr T(\lambda)\equiv
P(\lambda)$ yields Hamiltonians (integrals of motion) parametrizing the moduli
space ${\cal M}$ of the spectral curves, i.e. the moduli space of vacua of
physical theory. The replace $Y\equiv w-1/w$ transforms the curve
(\ref{scsl2m}) to the standard hyperelliptic form $Y^2=P^2-4$, the genus of
the curve being $N_c-1$.

As to the meromorphic 1-form, the remaining ingredient of the
SW anzatz, it is given by $dS=\lambda{dw\over
w}=\lambda {dP\over Y}$ and is just the shorten action "$pdq$"
along the non-contractible contours on the complex Lagrangian
tori. Its defining property is that the derivatives of
$dS$ with respect to the moduli (ramification points)
are holomorphic differentials on the spectral curve.

Note that the Toda chain which has two different Lax representations.
Indeed,
the system can be reformulated in terms of $N_c\times N_c$ matrix.
To this end,
consider the two-component Baker-Akhiezer function $\Psi_{n}=
\left(\begin{array}{c}
\psi_n\\
\chi_n
\end{array}\right)$.
Then the linear problem (\ref{lproblem}) can be rewritten as
\begin{equation}
\psi_{n+1}-p_n\psi_n+e^{q_n-q_{n-1}}\psi_{n-1}=\lambda\psi_n,\ \ \ \
\chi_n=-e^{q_{n-1}}\psi_{n-1}
\end{equation}
and, along with the periodic
boundary conditions (\ref{pbc}) reduces to the linear problem
${\cal L}(w)\Phi=\lambda\Phi$ for the $n\times n$ Lax operator
\begin{equation}
\label{LaxTC}
{\cal L}(w) =
\left(\begin{array}{ccccc}
-p_1 & e^{{1\over 2}(q_2-q_1)} & 0 & &
{1\over w}e^{{1\over 2}(q_{N_c}-q_1)}\\
e^{{1\over 2}(q_2-q_1)} & -p_2 & e^{{1\over 2}(q_3 - q_2)} & \ldots & 0\\
0 & e^{{1\over 2}(q_3-q_2)} & - p_3 & & 0 \\
 & & \ldots & & \\
{w}e^{{1\over 2}(q_{N_c}-q_1)} & 0 & 0 & & -p_{N_c}
\end{array} \right)
\end{equation}
with the $n$-component Baker-Akhiezer function
$\Phi=\{e^{-q_n/2}\psi_n\}$. This leads
us to the spectral curve
\begin{equation}\label{hren}
\det\left({\cal L}(w)-\lambda\right)=0
\end{equation}
which is still equivalent to the spectral curve (\ref{specurv}).

\paragraph{\large\bf Including fundamental matter.}
In order to include matter hypermultiplets, one needs to consider an
integrable system which generalizes the Toda chain. One may generalize both
$N_c\times N_c$ and $2\times 2$ Lax representations. The generalization of
the former one is well known to be the elliptic Calogero model \cite{Cal}. In
turn, now it is also known how to extend this system to other classical groups
\cite{HP}.

At the same time, in order to include the fundamental matter, one has to
generalize the $2\times 2$ representation of the Toda chain. This
generalization is the inhomogeneous XXX chain \cite{GMMM,GGM1}. The XXX spin
chain is given by the Lax operator
\be\label{LaxXXX}
L_i(\lambda) = (\lambda+\lambda_i)
\cdot {\bf 1} + \sum_{a=1}^3 S_{a,i}\cdot\sigma^a=
\left(
\begin{array}{cc}
\lambda+\lambda_i+S_{0,i}&S_{+,i}\\
S_{-,i}&\lambda+\lambda_i-S_{0,i}
\end{array}
\right)
\ee
Here $\lambda_i$'s are inhomogeneities , which can be
introduced since the Poisson brackets (\ref{quadr-r}) depends only on
difference of the spectral parameters.
The Lax operator (\ref{LaxXXX}) restores "the
symmetry" along the diagonal since has non-zero second diagonal entry linear
in the spectral parameter.  The Lax operator satisfies the Poisson algebra
(\ref{quadr-r}) with the same rational $r$-matrix (\ref{rat-r}).

The Poisson brackets of the dynamical variables $S_a$, $a=1,2,3$
(taking values in the algebra of functions)
are implied by (\ref{quadr-r}) and are just
\be\label{Scomrel}
\{S_a,S_b\} = -i\epsilon_{abc} S_c
\ee
i.e. $\{S_a\}$ plays the role of angular momentum (``classical spin'')
giving the name ``spin-chains'' to the whole class of systems.
Algebra (\ref{Scomrel}) has an obvious Casimir function
(an invariant, which Poisson commutes with all the spins $S_a$),
\be\label{Cas}
\left(C^{(2)}\right)^2 = {\bf S}^2 = \sum_{a=1}^3 S_aS_a
\ee

For the spin chain, the linear problem remains the same
(\ref{lproblem}) as well as the boundary conditions (\ref{pbc}). This means
that this spin chain is also associated with the Dynkin diagram of the
group $A_{N_c-1}^{(1)}$. The spectral curve now is of the general form
\be\label{scsl2m}
w^2-\Tr T(\lambda) w+\det T(\lambda)=
w^2-P(\lambda) w+Q(\lambda)=0
\ee
which also gets manifestly hyperelliptic (of the same genus $N_c-1$)
upon the replace $Y\equiv w-Q/w$:
\be
Y^2=P^2-4Q
\ee
The SW meromorphic 1-form also remains unchanged $dS=\lambda{dw\over
w}=\lambda{dP\over Y}$.

One can also define masses of the hypermultiplets immediately from
the spectral curve. They are just zeroes of the determinant of the
transfer matrix. Since
\be\label{detTxxx}
\det_{2\times 2} L_i(\lambda) = (\lambda+\lambda_i)^2 -
\left(C^{(2)}\right)^2
\ee
one gets
\be
\det_{2\times 2} T(\lambda) = \prod_{i=1}^{N_c}
\det_{2\times 2} L_i(\lambda) =
\prod_{i=1}^{N_c} \left((\lambda + \lambda _i)^2 -
\left(C^{(2)}_i\right)^2\right) = \\
= \prod_{i=1}^{N_c}(\lambda - m_i^+)(\lambda - m_i^-)
\ee
where we assumed that the values of spin $C^{(2)}$ can be different at
different sites of the chain, and
\be
m_i^{\pm} = -\lambda_i \pm C^{(2)}_i.
\label{mpm}
\ee

While determinant of the transfer matrix (\ref{detTxxx})
depends on dynamical variables
only through Casimirs $C^{(2)}_i$ of the Poisson algebra, dependence of
the trace $\Tr_{2\times 2}T(\lambda)$ is less trivial.
Still, it depends
on $S_a^{(i)}$ only through Hamiltonians of the spin chain (which are not
Casimirs but Poisson-commute with {\it each other}) -- see further details in
\cite{GMMM}.

Note that the above constructed system describes as many as $N_f=2N_c$
massive hypermultiplets. This corresponds to the "maximal" system
and is associated with the UV-finite theory.
This system depends on $N_c-1$ physical moduli (rank of the
group $SU(N_c)$) and on $2N_c$ masses. These moduli are parametrized
by the exactly $N_c-1$ independent integrals of motion\footnote{The $N_c$-th
integral of motion, the third projection of the full spin of the system
$\sum_i S_3^{(i)}$ is supposed to be zero. This condition is equivalent to
zero full momentum of the system in the Toda case above and is reflected in
absence of the $\lambda^{N_c-1}$-term in the polynomial $P(\lambda)$.}.
This system can be degenerated \cite{GGM1} to the theory with less number of
massive hypermultiplets by degenerating the Lax operators on several sites,
or to the pure gauge theory, with the Lax operators on all the sites being
"maximally" degenerated. In this latter case, we return to the Toda
chain \cite{GMMM}.

\paragraph{{\large\bf $D_n$ gauge group.}}
The main lesson of the previous section is that, in order to deal with
other gauge groups {\it and} fundamental matter, one needs to use the
$2\times 2$ Lax representation.
Now we turn to the first non-trivial example of the $D_n$ gauge group.
We work out this example in some detail and further make shorter comments
for the other groups. As above, we begin with the
$D_n^{(1)}$-Toda chain and then generalize it to the spin chain.

The Toda chain associated with the $D_{n}^{(1)}$ group first was applied to
the description of the SW anzatz for the $SO(2n)$ pure gauge theory in
\cite{MW}. This system can be also described by two different Lax
representations. First of all, there is $2n\times 2n$ representation
\cite{BOP}
\be
{\cal L}(w)=\left(
\begin{array}{cc}
l&c\\
-\bar c&-l
\end{array}
\right)
\ee
where $l$, $c$ and $\bar c$ are $n\times n$ matrices
\be
\left(l\right)_{ij}=\delta_{ij}p_j+\left(\delta_{j,k-1}
e^{{1\over 2}(q_j-q_{j+1})} +\delta_{j-1,k}e^{{1\over 2}(q_k-q_{k+1})}
\right),\\
\left(c\right)_{ij}=(\delta_{j1}\delta_{k2} +\delta_{j2}\delta_{k1})
e^{-{1\over 2}(q_1+q_2)}+w
(\delta_{j,n-1}\delta_{kn} +\delta_{jn}\delta_{k,n-1})
e^{{1\over 2}(q_{n-1}+q_n)},\\
\left(\bar c\right)_{ij}=(\delta_{j1}\delta_{k2} +\delta_{j2}\delta_{k1})
e^{-{1\over 2}(q_1+q_2)}+{1\over w}
(\delta_{j,n-1}\delta_{kn} +\delta_{jn}\delta_{k,n-1})
e^{{1\over 2}(q_{n-1}+q_n)}
\ee
One gets for the linear problem with such Lax operator ${\cal
L}(w)\Phi=\lambda\Phi$ the spectral curve (\ref{hren}) for the
$D_{n}^{(1)}$-Toda (with replace $w\to w/\lambda^2$)
\be\label{DTodasc}
w^2-P_{n}(\lambda^2)w+\lambda^4=0
\ee

Now we show how this Toda can be described within the framework of the
$2\times 2$ Lax matrices. The idea of this description goes back to
E.Sklyanin \cite{Skl88}, who proposed to change boundary conditions of the
usual spin chain in accordance with the Dynkin diagram of the group
$D_n^{(1)}$, see Fig.2.


\begin{figure}[t]
\special{em:linewidth 0.4pt}
\unitlength 1.00mm
\linethickness{0.4pt}
\begin{picture}(96.67,90.00)
\put(29.00,80.00){\circle{2.00}}
\put(43.00,80.00){\circle{2.00}}
\put(54.00,80.00){\circle{0.67}}
\put(58.00,80.00){\circle{0.67}}
\put(62.67,80.00){\circle{0.67}}
\put(73.67,80.00){\circle{2.00}}
\put(85.33,80.00){\circle{2.00}}
\emline{30.33}{80.00}{1}{42.00}{80.00}{2}
\emline{75.00}{80.00}{3}{84.33}{80.00}{4}
\put(58.00,62.66){\makebox(0,0)[cc]{Fig.2 Dynkin diagram for $D_n^{(1)}$}}
\put(18.33,89.00){\circle{2.00}}
\put(18.33,68.67){\circle{2.00}}
\put(95.67,68.67){\circle{2.00}}
\put(95.67,89.00){\circle{2.00}}
\emline{19.33}{88.33}{5}{28.00}{80.67}{6}
\emline{28.33}{79.33}{7}{19.00}{69.33}{8}
\emline{86.33}{80.67}{9}{94.67}{88.33}{10}
\emline{86.00}{79.33}{11}{95.00}{69.33}{12}
\end{picture}
\vspace{-5.7cm}
\end{figure}

The rough description of the procedure looks as
follows (see Fig.3). First, one presents the closed diagram, Fig.3a, with two
unusual Lax operators $K_+$ and $K_-$ inserted. When constructing the
transfer-matrix, this diagram is passed around
\be
T=K_+^{(1)}L_3\ldots L_{n}K_-^{(2)}{\bar L}_{n}\ldots {\bar L}_3
\ee
Now let us identify $L_i(\lambda)$ with ${\bar L}_i^{-1}(-\lambda)$. This
means gluing the upper and lower vertices of the Dynkin diagram together. It
gives the diagram of the form Fig.3b.

\begin{figure}[h]
\vspace{1cm}
\special{em:linewidth 0.4pt}
\unitlength 1.00mm
\linethickness{0.4pt}
\begin{picture}(145.01,95.00)
\emline{17.33}{80.00}{1}{28.00}{80.00}{2}
\put(29.00,80.00){\circle{2.00}}
\put(43.00,80.00){\circle{2.00}}
\put(54.00,80.00){\circle{0.67}}
\put(58.00,80.00){\circle{0.67}}
\put(62.67,80.00){\circle{0.67}}
\put(73.67,80.00){\circle{2.00}}
\put(85.33,80.00){\circle{2.00}}
\emline{30.33}{80.00}{3}{42.00}{80.00}{4}
\emline{75.00}{80.00}{5}{84.33}{80.00}{6}
\emline{86.33}{80.00}{7}{97.33}{80.00}{8}
\put(124.66,95.00){\makebox(0,0)[cc]{(a)}}
\emline{17.00}{80.00}{9}{14.67}{80.00}{10}
\put(102.33,83.67){\circle{11.35}}
\put(8.33,83.33){\circle{11.10}}
\emline{14.33}{80.00}{11}{13.00}{80.00}{12}
\put(8.00,83.00){\makebox(0,0)[cc]{$K_+^{(1)}$}}
\put(102.33,83.00){\makebox(0,0)[cc]{$K_-^{(2)}$}}
\emline{17.66}{86.33}{13}{28.33}{86.33}{14}
\put(29.33,86.33){\circle{2.00}}
\put(43.33,86.33){\circle{2.00}}
\put(54.33,86.33){\circle{0.67}}
\put(58.33,86.33){\circle{0.67}}
\put(63.00,86.33){\circle{0.67}}
\put(74.00,86.33){\circle{2.00}}
\put(85.66,86.33){\circle{2.00}}
\emline{30.66}{86.33}{15}{42.33}{86.33}{16}
\emline{75.33}{86.33}{17}{84.66}{86.33}{18}
\emline{17.33}{86.33}{19}{15.00}{86.33}{20}
\emline{14.67}{86.33}{21}{13.33}{86.33}{22}
\emline{86.67}{86.33}{23}{97.33}{86.33}{24}
\put(66.67,76.00){\vector(1,0){0.2}}
\emline{50.00}{76.00}{25}{66.67}{76.00}{26}
\put(50.00,89.33){\vector(-1,0){0.2}}
\emline{66.67}{89.33}{27}{50.00}{89.33}{28}
\put(124.66,34.00){\makebox(0,0)[cc]{(b)}}
\put(102.33,48.00){\circle{11.35}}
\put(8.33,47.67){\circle{11.10}}
\put(8.00,47.33){\makebox(0,0)[cc]{$K_+^{(1)}$}}
\put(102.33,47.33){\makebox(0,0)[cc]{$K_-^{(2)}$}}
\put(29.33,47.33){\circle{2.00}}
\put(43.33,47.33){\circle{2.00}}
\put(54.33,47.33){\circle{0.67}}
\put(58.33,47.33){\circle{0.67}}
\put(63.00,47.33){\circle{0.67}}
\put(74.00,47.33){\circle{2.00}}
\put(85.66,47.33){\circle{2.00}}
\emline{30.66}{47.33}{29}{42.33}{47.33}{30}
\emline{75.33}{47.33}{31}{84.66}{47.33}{32}
\emline{12.67}{51.67}{33}{28.33}{47.33}{34}
\emline{28.33}{47.33}{35}{12.67}{43.67}{36}
\emline{98.00}{52.00}{37}{87.00}{47.33}{38}
\emline{87.00}{47.33}{39}{99.00}{43.33}{40}
\put(58.00,22.33){\makebox(0,0)[cc]
{Fig.3 Obtaining $D^{(1)}_{n}$ from the ring diagram of the
$A^{(1)}$-type}}
\put(58.33,55.33){\vector(0,-1){0.2}}
\emline{58.33}{69.00}{41}{58.33}{55.33}{42}
\put(112.67,64.33){\circle{12.07}}
\put(112.67,64.00){\makebox(0,0)[cc]{$K$}}
\put(125.67,64.00){\makebox(0,0)[cc]{=}}
\put(139.00,64.33){\circle{12.02}}
\put(139.00,67.00){\circle{2.11}}
\put(139.00,61.33){\circle{2.00}}
\emline{138.00}{61.33}{43}{129.33}{61.33}{44}
\emline{138.00}{67.00}{45}{129.67}{67.00}{46}
\emline{107.33}{67.00}{47}{102.00}{67.00}{48}
\emline{107.67}{61.33}{49}{102.33}{61.33}{50}
\end{picture}
\vspace{-1.5cm}
\end{figure}

Although this procedure may seem too pictorial, it turned out to be correct.
Indeed, let us consider a system with the transfer matrix
of the form
\be
T(\lambda)\equiv K_+(\lambda)K_-(\lambda)
\ee
and discuss properties of the reflection matrices $K_{\pm}$.
To keep the system integrable (i.e. traces
of the transfer-matrices commuting),
one should impose {\it the reflection equation}
\be\label{RE}
\left\{\stackrel{_{_1}}{K}_{\pm}(\lambda),
\stackrel{_{_2}}{K}_{\pm}(\lambda')\right\}=
\left[r(\lambda -\lambda'),
\stackrel{_{_1}}{K}_{\pm}(\lambda)
\stackrel{_{_2}}{K}_{\pm}(\lambda') \right]+\\+
\stackrel{_{_1}}{K}_{\pm}(\lambda)
r(\lambda +\lambda')
\stackrel{_{_2}}{K}_{\pm}(\lambda')-
\stackrel{_{_2}}{K}_{\pm}(\lambda')
r(\lambda +\lambda')
\stackrel{_{_1}}{K}_{\pm}(\lambda)
\ee
where $\stackrel{_{_1}}{K}_{\pm}(\lambda)\equiv
{K}_{\pm}(\lambda)\otimes {\bf 1}$,
$\stackrel{_{_2}}{K}_{\pm}(\lambda)\equiv
{\bf 1}\otimes {K}_{\pm}(\lambda)$, $K_-(\lambda)=K^t_+(-\lambda)$ --
transposed $K_+$-matrix and $r$-matrix being that of the corresponding
dynamical system.

Now one can note that the product
$L(\lambda)K_{\pm}(\lambda)L^{-1}(-\lambda)$
satisfies the reflection equation
(\ref{RE}) whenever $K$ satisfies it, and $L$ is the standard Lax
operator that satisfies the Yang-Baxter equation (\ref{quadr-r}).
Therefore, the expression
\be\label{T}
T(\lambda)=K_+^{(1)}(\lambda)L_3(\lambda)\ldots L_{n}
(\lambda)K_-^{(2)}(\lambda) {L}_{n}^{-1}(-\lambda)
\ldots {L}_3^{-1}(-\lambda)
\ee
is, indeed, a good candidate for the transfer matrix in the system described
by the picture of Fig.3.

In order to find particular solutions $K_{\pm}(\lambda)$
to the reflection equation describing the $D_n^{(1)}$-Toda chain,
take into account that they are associated,
in accordance with Fig.2, with the pairs of vertices at each "end"
of the Dynkin diagram respectively. This
hints to look them for as quadratic polynomial ($2\times 2$) matrices.
The Toda chain solution of such a type can be found from the commutation
relations (\ref{RE}). Indeed, one can construct the solution in the standard
way: first, consider the general $2\times 2$ matrix $K(\lambda)$

\be\label{K}
K(\lambda)=\left(
\begin{array}{cc}
A(\lambda)&B(\lambda)\\
C(\lambda)&D(\lambda)
\end{array}
\right)
\ee
with all entries being quadratic polynomials (this concrete
form can be achieved by
the proper shift of $\lambda$)
\be\label{ABCD}
A(u)=\alpha \lambda^2 +A_{1}\lambda +A_{0} \\
B(u)=\beta \lambda^2 +B_{0} \\
C(u)=\gamma \lambda^{2} +C_{0}\\
D(\lambda)=-A(-\lambda)
\ee
Such a matrix celebrates the property $K(\lambda)=K^{-1}(-\lambda)$.
Inserting $K(\lambda)$ into (\ref{RE}), one immediately
gets \cite{Kuz2} the quadratic Poisson algebra
\be\label{pb}
\left\{A_{0},A_{1}\right\}=\beta C_{0}-\gamma B_{0},\ \ \
\left\{C_0,A_{1}\right\}= 2\gamma A_{0}-2\alpha C_{0},\ \ \
\left\{B_{0},A_{1}\right\}= 2\alpha B_{0}-2\beta A_0,\\
\left\{B_{0},A_{0}\right\}=2A_1 B_{0},\ \ \
\left\{C_{0},A_{0}\right\}= -2A_{1}C_{0},\ \ \
\left\{C_{0},B_{0}\right\}= 4A_{1}A_{0}
\ee
with $\alpha$, $\beta$ and $\gamma$ being central elements of the algebra
(\ref{pb}) (any one of such (non-zero) elements can be reduced to unity by
rescaling $\lambda$). The determinant of the reflection matrix is equal to
\be
\det K(\lambda)= -(\alpha^2 +\beta\gamma)\lambda^4 +
Q_{2}\lambda^2 -Q_{0}
\ee
where the Casimir elements $Q_{0}$, $Q_{2}$ are
\be
Q_{2}=A_{1}^2 -2\alpha A_{0} -\beta C_{0} -\gamma B_{0}, \ \ \ \
Q_{0}=A_{0}^2 +B_{0}C_{0}.
\ee
These solutions in generic case
can be interpreted as a pair of $SO(3)$ or $SO(2,1)$ attached to
one cite.

Now one can note that this solution gives too general curve -- for the Toda
chain one does not need any free parameters (Casimir elements) associated
with the ending cites. Indeed, the suitable Toda solution for
$K_-(\lambda)$ can be associated with special values of the Casimir and
central elements $Q_0=0$, $Q_2=1$, $\alpha=0$, $\gamma=0$ and $\beta=1$
(isomorphic case is $\beta=0$ while $\gamma =1$ associated with
$K_+(\lambda)$) \cite{Kuz1}
\be
K_-(\lambda)=\left(
\begin{array}{cc}
A_1\lambda+A_0&\lambda^2+B_0\\
C_0&A_1\lambda-A_0
\end{array}
\right)
\ee
In this case, the Poisson algebra (\ref{pb})
can be bosonized $A_1=e^q$, $A_0=2pe^q$, $C_0=2e^q\sinh q$ and
$B_0=-2p^2e^{-q} \sinh q$ with $\{p,q\}=1$.

Then, the spectral curve can be easily obtained from (\ref{specurv}), i.e.
from
\be
w^2-\Tr T\cdot w+\det T=0
\ee
with
the transfer matrix from (\ref{T}) and using the manifest formulas (\ref{K})
and (\ref{ABCD}). The result is exactly (\ref{DTodasc}), and can be obtained
using that $\det T(\lambda)=\det K_+^{(1)}\det K_-^{(2)}=\lambda^4$ and the
following

\paragraph{Lemma.}
For
arbitrary $2\times 2$-matrix $M(\lambda)$ with polynomial entries such that
$M_{11}(\lambda)=(-)^\epsilon M_{22}(-\lambda)$ ($\epsilon=0,1$) and
$M_{12}(\lambda)$, $M_{21}(\lambda)$ are of a certain (the same) parity, the
product $L^{-1}(-\lambda)M(\lambda)L(\lambda)$ with $L(\lambda)$ from
(\ref{TodaLax}) or (\ref{LaxXXX}) possesses the same properties.

\medskip

Similarly to the $SU(N_c)$ case, in order to include the fundamental matter,
one just needs to the symmetry (this time, of the other diagonal (12),(21))
which releases the constraints on Casimir elements of the
reflection matrices. Certainly, one also needs to
substitute the Toda Lax operators by the
spin chain operators
(\ref{LaxXXX}) (the resulting system is sometimes called the spin chain
interacting with the tops at the ends).
Thus,
this time one imposes onto the reflection matrix
the only condition $Q_0=0$ (with
$\alpha=0$, $\beta=\gamma=1$)\footnote{Note that, in variance with
the spin chain Lax operator, one can not
shift $\lambda$ by a constant inhomogeneity in this matrix since
the reflection equation (\ref{RE}) depends not only on difference of the
spectral parameters.}
\be\label{aa}
K_-(\lambda)=\left(
\begin{array}{cc}
A_1\lambda+A_0&\lambda^2+B_0\\
\lambda^2+C_0&A_1\lambda-A_0
\end{array}
\right),\ \ \ \ \ A_0^2+B_0C_0=-Q_0=0
\ee
Then, the determinant of the corresponding
transfer matrix immediately leads
to the spectral curve \cite{AS} (taking into account the lemma)
\be\label{scD}
w^2 - P_{n}(\lambda^2)w +\lambda^4 Q(\lambda^2)=0
\ee
where $P_{n}(\lambda^2)=
\prod^{n}(\lambda^2-\lambda_i^2)$ and
$\lambda_i$'s give $n$ independent
gauge moduli (=Hamiltonians of the spin chain).
This number $n$ is exactly the number of sites (counting $K_{\pm}$) on
the diagram of Fig.3b.
Zeroes of the polynomial $Q(\lambda)$ give masses of the
hypermultiplets
\be
Q(\lambda^2)=\left(\lambda^2-Q_2^{(1)}\right)
\left(\lambda^2-Q_2^{(2)}\right)
\prod_i^{n-2} \left(\lambda^2 - (m^+_i)^2\right)
\left(\lambda^2 - (m^-_i)^2\right)
\ee
with $m^{\pm}_i$ as in (\ref{mpm}).
Note that four of the masses are associated with the Casimir functions
corresponding to the reflection matrices, not to the Lax operators. This
means that, in contrast to the $SU(N_c)$ case, the four masses are
distinguished. We return to this issue again in the paragraph devoted to the
brane picture.

Now let us note that the number of hypermultiplets described by the curve
(\ref{scD}) is $2n-2$. Similarly to the $SU(N_c)$ case, it should be
associated with the "maximal" UV-finite system. This is, indeed, the case,
since the $\beta$-function for the SQCD with $SO(2n)$ gauge
group corresponding to the $D_n^{(1)}$ system and $N_f$
fundamental hypermultiplets is equal to zero when $N_f=2n-2$. Note that this
system is described by $2n$ gauge moduli $\pm\lambda_i$, in accordance with
(\ref{scD}).

\paragraph{{\large\bf $B_n$ and $C_n$ gauge groups.}}
Now we are going to consider the remaining classical groups. We start with the
symplectic group. In accordance with the general rule \cite{MW}, in order
to describe theory with $Sp(2n)$ gauge group, one needs to
consider the dual group $\left(C_n^{(1)}\right)^{\vee}$. The corresponding
Dynkin diagram is depicted in Fig.4. In accordance with this diagram,
we define the transfer matrix
\be\label{TC}
T(\lambda)=K_+^{(1)}(\lambda)L_3(\lambda)\ldots L_{n+2}
(\lambda)K_-^{(2)}(\lambda) {L}_{n+2}^{-1}(-\lambda)
\ldots {L}_3^{-1}(-\lambda)
\ee
where the reflection matrices now correspond to single-root ending sites
with two outcoming lines each, instead of two-root end as in Fig.3.

\begin{figure}[t]
\special{em:linewidth 0.4pt}
\unitlength 1.00mm
\linethickness{0.4pt}
\begin{picture}(99.00,82.67)
\put(29.00,80.00){\circle{2.00}}
\put(43.00,80.00){\circle{2.00}}
\put(54.00,80.00){\circle{0.67}}
\put(58.00,80.00){\circle{0.67}}
\put(62.67,80.00){\circle{0.67}}
\put(73.67,80.00){\circle{2.00}}
\put(85.33,80.00){\circle{2.00}}
\emline{30.33}{80.00}{1}{42.00}{80.00}{2}
\emline{75.00}{80.00}{3}{84.33}{80.00}{4}
\put(58.00,62.66){\makebox(0,0)[cc]{Fig.5 Dynkin diagram for
$\left(C_n^{(1)}\right)^\vee=D^{(2)}_{n+1}$}}
\put(98.00,80.00){\circle{2.00}}
\put(16.33,80.00){\circle{2.00}}
\emline{17.33}{80.67}{5}{28.00}{80.67}{6}
\emline{17.33}{79.00}{7}{28.67}{79.00}{8}
\emline{86.33}{80.67}{9}{97.33}{80.67}{10}
\emline{86.00}{79.00}{11}{97.67}{79.00}{12}
\emline{90.33}{82.67}{13}{93.33}{80.00}{14}
\emline{93.33}{80.00}{15}{90.33}{77.33}{16}
\emline{20.67}{80.00}{17}{24.00}{82.67}{18}
\emline{20.67}{80.00}{19}{23.33}{77.33}{20}
\end{picture}
\vspace{-5.5cm}
\end{figure}

Therefore, it is natural this time to look for the
reflection matrices linear in $\lambda$. We can apply the same approach as
above in order to obtain the solution \cite{Skl88,Kuz2}
\be\label{Klinear}
K(\lambda)=\left(
\begin{array}{cc}
\alpha \lambda + A_0&\beta\lambda\\
\gamma\lambda&-\alpha\lambda+A_0
\end{array}
\right)
\ee
In variance with the quadratic reflection matrix, the entries of this
solution to the reflection equation are all Poisson commuting, i.e. the
reflection matrix (\ref{Klinear}) does not contain dynamical variables.

In order to describe the Toda chain, as usual one has to
restrict this reflection matrix requiring $\alpha=\gamma=0$, $\beta=1$ for
$K_-(\lambda)$ and $\det K(\lambda)\equiv Q_2=A_0^2=1$
\be
K(\lambda)=\left(
\begin{array}{cc}
1&\lambda\\
0&1
\end{array}
\right)
\ee
Including the
fundamental matter implies, as above, restoring of the symmetry of this
matrix, with the arbitrary $Q_2=A_0^2$:
\be\label{refC}
K(\lambda)=\left(
\begin{array}{cc}
A_0&\lambda\\
\lambda&A_0
\end{array}
\right)
\ee

To obtain the spectral curve, we need to use the
lemma, and also to note that $\Tr T(0)=2$ in the Toda case and $\Tr T(0)=
2\sqrt{Q_2^{(1)}Q^{(2)}_2}
\prod_i m^+_i m^-_i$. This latter fact is
trivially obtained since K is diagonal. Then, one obtains for the Toda
spectral curve \cite{Kuz3,MW} (remark the misprint in \cite{MW})\footnote{As
before, this spectral curve can be also immediately presented as
(\ref{hren}) with the $2(n+1)\times 2(n+1)$ Lax operator \cite{BOP}.
}
\be
w^2 -\left(\lambda^2 P_{n}(\lambda^2)-2\right)\cdot w+1=0
\ee
and for the spin chain spectral curve \cite{AS}
\be\label{scC}
w^2 -\left(\lambda^2 P_{n}(\lambda^2)-
2\sqrt{Q_2^{(1)}Q^{(2)}_2}
\prod_i m^+_i m^-_i\right)\cdot w +Q(\lambda)=0
\ee
where, as before, polynomial $P_{n}(\lambda^2)$ depends on
$n$ independent gauge moduli and
zeroes of the polynomial $Q(\lambda)$ give masses of the
hypermultiplets
\be
Q(\lambda^2)=\left(\lambda^2-Q_2^{(1)}\right)
\left(\lambda^2-Q_2^{(2)}\right)
\prod_i^{n} \left(\lambda^2 - (m^+_i)^2\right)
\left(\lambda^2 - (m^-_i)^2\right)
\ee
with $m^{\pm}_i$ as in (\ref{mpm}).

Note that this time the number of independent gauge moduli, i.e. of
Hamiltonians in integrable system is equal to the number of internal sites
of the Dynkin diagram, Fig.4. This reflects the fact the ending ($K$-)
matrices do not contain dynamical degrees of freedom.

Now we come to the last series of the classical groups -- the odd orthogonal
series. The Dynkin diagram of the group $\left(B_n^{(1)}\right)^{\vee}$
describing the theory with the $SO(2n+1)$ gauge group is
given in Fig.5 and the corresponding transfer matrix is
\be\label{TB}
T(\lambda)=K_+^{(1)}(\lambda)L_3(\lambda)\ldots L_{n+1}
(\lambda)K_-^{(2)}(\lambda) {L}_{n+1}^{-1}(-\lambda)
\ldots {L}_3^{-1}(-\lambda)
\ee

\begin{figure}[t]
\special{em:linewidth 0.4pt}
\unitlength 1.00mm
\linethickness{0.4pt}
\begin{picture}(100.33,90.00)
\put(29.00,80.00){\circle{2.00}}
\put(43.00,80.00){\circle{2.00}}
\put(54.00,80.00){\circle{0.67}}
\put(58.00,80.00){\circle{0.67}}
\put(62.67,80.00){\circle{0.67}}
\put(73.67,80.00){\circle{2.00}}
\put(85.33,80.00){\circle{2.00}}
\emline{30.33}{80.00}{1}{42.00}{80.00}{2}
\emline{75.00}{80.00}{3}{84.33}{80.00}{4}
\put(58.00,62.66){\makebox(0,0)[cc]{Fig.5 Dynkin diagram for
$\left(B_n^{(1)}\right)^\vee=A^{(2)}_{2n-1}$}}
\put(18.33,89.00){\circle{2.00}}
\put(18.33,68.67){\circle{2.00}}
\emline{19.33}{88.33}{5}{28.00}{80.67}{6}
\emline{28.33}{79.33}{7}{19.00}{69.33}{8}
\put(99.33,80.00){\circle{2.00}}
\emline{86.33}{80.67}{9}{98.33}{80.67}{10}
\emline{86.00}{79.00}{11}{98.67}{79.00}{12}
\emline{93.00}{82.67}{13}{88.67}{80.00}{14}
\emline{88.67}{80.00}{15}{92.67}{77.33}{16}
\end{picture}
\vspace{-5.5cm}
\end{figure}

From our previous consideration we know the reflection
matrices that are to be associated with the ends of this Dynkin diagram --
one matrix is quadratic (\ref{K})-(\ref{ABCD}),
the other one is linear. Note, however, that, in contrast to the symplectic
case, the arrow at the end of the diagram is directed inside. Therefore,
although being linear, the corresponding reflection matrix is slightly
modified as compared with (\ref{Klinear}). Namely, in accordance with
\cite{Skl88,Kuz3} it should be chosen for the Toda chain in the form
\be\label{KlinearB}
K(\lambda)=\left(
\begin{array}{cc}
0&-{1\over 2}\lambda\\
2\lambda&0
\end{array}
\right)
\ee
Coefficient 2 here can be replaced by any non-unit number not changing
the final result. As before,
we can use the lemma in order to derive the spectral curves. For the
Toda chain it is \cite{Kuz3,MW}\footnote{Again,
this spectral curve is obtained from (\ref{hren}) related to the
linear problem for the $2n\times 2n$
Lax operator \cite{BOP}. To be absolutely precise, it comes with
$w\to \lambda w$ in (\ref{BToda}).}
\be\label{BToda}
w^2-\lambda P_{n}(\lambda^2)\cdot w+\lambda^4=0
\ee

In variance with the previous cases, the Toda reflection matrix
(\ref{KlinearB}) is of quite symmetric form and, therefore,
one should not change it in
order to include the fundamental matter. The only thing to be done is
that, instead of
coefficient 2 in (\ref{KlinearB}) there should
be introduced an arbitrary coefficient $\rho\ne 1$ that plays the role of the
only Casimir element (this coefficient becomes essential in the spin chain
case, see \cite{GGM1}).

Certainly, the second,
quadratic reflection matrix is changed as for the $D$ series
(\ref{aa}). Then, one
easily gets for the spin chain spectral curve \cite{AS}
\be\label{scB}
w^2-\lambda P_{n}(\lambda^2)\cdot w+\lambda^4 Q(\lambda)=0
\ee
with
\be
Q(\lambda^2)\sim\left(\lambda^2-Q_2^{(1)}\right)
\prod_i^{n-1} \left(\lambda^2 - (m^+_i)^2\right)
\left(\lambda^2 - (m^-_i)^2\right)
\ee
and $m^{\pm}_i$ as in (\ref{mpm}). In this case the $n$ independent gauge
moduli are associated with the internal sites of the Dynkin diagram, Fig.5
and with the left reflection matrix depending on dynamical variables.

Note that the property of obtaining "maximal" UV finite system for the spin
chain case is preserved for all the classical groups. This property is
equivalent to impossibility arbitrary change the ratio of normalizations
$P(\lambda)=\Tr T$ and $Q(\lambda)=\det T$. In turn, this ratio is changed by
twisting procedure \cite{GGM1} and can be rescaled for any
degenerated case similar to the Toda chain -- see discussion in
\cite{GGM1}.

\paragraph{{\large\bf Brane picture.}}

An alternative way to obtain the curves for SUSY YM theory
is to construct the brane picture behind it. The summary of
the brane/integrability correspondence can be found in \cite{ag}.
In accordance with the standard type IIA picture applicable to
description of the theory with $SU(N_c)$ gauge group and $N_f=2N_c$
fundamental hypermultiplets\footnote{We discuss here only this maximally
non-degenerated case. Degenerations are immediate.} \cite{W}, there are two
parallel NS5 branes filling the whole volume in
$(x^0,x^1,x^2,x^3,x^4,x^5)$-directions, with $N_c$ D4 branes stretched
between them in $x^6$ direction. These D4 branes lie along
$(x^0,x^1,x^2,x^3,x^6)$ coordinates and can move in the
$(x^4,x^5)$ plane giving $N_c-1$ moduli (the center of masses
is decoupled corresponding to the $U(1)$ factor).

This picture is associated with the pure gauge theory. One can include the
fundamental matter hypermultiplets in two possible ways. The first one is to
add $N_f$ semi-infinite D4 branes ending on the NS5 branes. This
corresponds to the $N\times N$ description of integrable system
\cite{GGM1}. The second way associated with the $2\times 2$ Lax
representation is to consider instead of semi-infinite branes D6 branes
extended in $(x^0,x^1,x^2,x^3,x^7,x^8,x^9)$ directions. The spin
chain description implies that with each D4 brane one associates a pair of
D6 branes (inducing hypermultiplet masses (\ref{mpm})), which gives in total
$N_f=2N_c$ (degenerations mean decoupling of some D6 branes). Positions of
the D6 branes in $(x^4,x^5)$ plane give masses of the corresponding
hypermultiplets.

In order to generalize this picture to other classical groups, one should
introduce into play an orientifold. This may be either O4
orientifold plane \cite{orien}, or O6 orientifold \cite{O6}\footnote{In
fact, the two cases are not absolutely equivalent:
using different orientifolds gives rise to different spectral curves
coinciding only after a redefinition of the spectral parameter
$w$. In particular, in the spectral curve equation for the
orthogonal group (\ref{scD}) there naturally emerges $\lambda^2 w$ for O4
\cite{orien} instead of $w$ for O6. These different spectral parameters
appear in different, respectively $N\times N$ and $2\times 2$ Lax
representations (cf. \cite{GGM1}). Therefore, one can identify O6 orientifold
with $2\times 2$ representation and, thus, with the spin chain.}. The
role of orientifold is to project out the brane picture to that with all
branes having their mirror images with respect to the orientifold.  Then,
each "physical" brane is associated with a pair of mirror branes. This
procedure of gluing mirror images is much similar to the picture of Fig.3
and, in fact, just corresponds to description of a classical group $G$ as
${\bf Z}_2$-embedding into $SU(N)$ group of enough higher $N$ (roughly,
$N=2\times$rank of $G$).

The Op orientifold has charge $\pm 2^{p-4}$ in units of the Dp brane charge.
Say, O4 orientifold plane has charge $\pm 1$ of the D4 charge. This
orientifold fills the same volume as D4 branes and is placed at $x^4=x^5=0$.
Different signs of the orientifold charge are assigned with orientifold
projecting onto different groups. In fact, the charge assignment is quite
tricky \cite{orien} -- (+,-,+) for the three regions to the left, between and
to the right of the NS5 branes for even orthogonal groups, (+,0,+) for odd
orthogonal groups and (-,+,-) for symplectic groups. Further details on the
brane picture may be found in \cite{orien}.

Similarly, the O6 orientifold has the world-volume extended in
$(x^0,x^1,x^2,x^3,x^7,x^8,x^9)$ directions and the charge $\pm 4$ of the D6
charge. This time the sign assignment is much simpler since the orientifold
looks not like a line, but like a point in the essential
($x^4,x^5,x^6$)-space.  Therefore, there is no room for several regions with
different signs. In fact, the sign "+" is associated with the orientifold
projection onto orthogonal groups on the world-volume of D4
branes\footnote{This sign corresponds to symplectic group on D6 branes. Since
this system describes the flavour group and because of the dulaity between
flavour and gauge group, one can conclude that symplectic groups on D6 branes
correspond to orthogonal (gauge) groups on D4 branes \cite{O6}.}, while the
sign "-" -- with the projection onto symplectic groups. In contrast with the
O4 orientifold, the whole picture with this O6 orientifold looks quite
immediately related to the integrable spin chains described in the paper, and
we are going to make some comments here on the detailed picture, following
mostly \cite{O6}.

Let us first describe the $SO(2n)$ case. In this case, in accordance with
Fig.3, we have $n-2$ physical D4 branes (or $2(n-2)$ D4 branes and their
mirror images) giving $n-2$ independent gauge moduli\footnote{Note
that in presence of the
orientifold there is no decoupling of the center of masses (i.e. of the
$U(1)$ factor).} (since brane
mirror pairs moves dependently) and associated with the internal sites of the
Dynkin diagram and a pair of (physical) D6 branes associated with each of
these D4 branes. The mirror structure is reflected in the form of polynomials
$P(\lambda^2)$ and $Q(\lambda^2)$ in (\ref{scD}) which have zeroes at
$\pm\lambda_i$ and $\pm m_i$ respectively.

Besides, there are two mirror pairs of D4 branes (or two physical
D4 branes) associated immediately with the orientifold, which correspond to
the reflection matrices. These 4 (2 physical) branes can move, since the
reflection matrices depend on dynamical variables, and, therefore,
correspond to two more independent gauge moduli. Each of these distinguished
branes is typically assigned with a pair of D6 branes. While one D6 brane of
the pair has arbitrary $(x^4,x^5)$ coordinates which give the mass of the
corresponding hypermultiplet, the other brane of the pair should be fixed in
$x^4=x^5=0$. This pair of D6 branes has the mirror pair assigned with the
mirror D4 brane. Totally, one has 4 coinciding D6 branes at $x^4=x^5=0$ which
contribute the factor $\lambda^4$ into the last term of (\ref{scD}) and
corresponds to the charge +4 of the orientifold O6.  Therefore, the
orientifold in this picture looks as consisting of 4 coinciding D6 branes.
Similarity between orientifold and several coinciding branes has
been already discussed in many places (in the very close context, in
\cite{Ur}).

This picture can be almost literally transformed to the $SO(2n+1)$ group. The
only change in this case is that one should associate with one of the
reflection matrices (the right one in Fig.5) the D4 brane placed in
$x^4=x^5=0$. Therefore, there are no gauge moduli associated with this brane
(it can not move) and one should consider polynomial $\lambda P(\lambda^2)$
in (\ref{scB}) instead of $P(\lambda^2)$ in (\ref{scD}). Besides, this brane
has no mirror since intersects with the orientifold in $(x^4,x^5,x^6)$-space.
Therefore, there are only two D6 branes associated with the corresponding end
of the Dynkin diagram, both with coordinates $x^4=x^5=0$. Thus, it still
produces $\lambda^4$-term and the four coinciding D6 branes consisting
the orientifold, but this time there are no hypermultiplets assigned with
this end of the Dynkin diagram. This is reflected in the form of polynomial
$Q(\lambda^2)$ (\ref{scB}).

An analogous description for the symplectic group looks far more tricky
\cite{O6}. Not entering the details, let us note that this time one should
reproduce the orientifold charge -4.
Therefore, it can not be produced
merely by 4 D6 branes. Instead, one could use
some "anti-D6-branes" associated with some "anti-D4-brane", each of them
contributing $\lambda^{-1}$ instead of $\lambda$. Then, one places
these "anti-D4-branes" onto the axis $x^4=x^5=0$ (therefore,
there are no gauge moduli associated with them). Each of them is accompanied
by a pair of mirror "anti-D6-branes" placed in $\pm {1\over \sqrt{Q_2^{(i)}}}$
respectively. This leads to the curve
\be\label{scC'}
w^2 -\left(P_{n}(\lambda^2)-
2\sqrt{Q_2^{(1)}Q^{(2)}_2}
\prod_i m^+_i m^-_i{1\over \lambda^2}\right)\cdot w +{Q(\lambda)\over
\lambda^4}=0
\ee
the term $\lambda^{-2}$ being due to the contributions of the
"anti-D4-branes". The constant in front of it can be fixed from requirement
that the expression for the curve is to be a perfect square at $\lambda=0$
\cite{Ur}.
This curve can be reduced to (\ref{scC}) by the replace $w\lambda^2\to w$.
There is also a different brane picture in this case (see, e.g.,
\cite{Ur}) leading directly to the curve (\ref{scC}).

Note that one would obtain the spectral curve of the spin chain in the form
(\ref{scC'}) choosing different normalization of the reflection matrix
(\ref{refC}) (this normalization makes some sense, see \cite{Kuz2})
\be\label{refC'}
K(\lambda)=\left(
\begin{array}{cc}
{1\over \lambda}&B_0\\
B_0&{1\over\lambda}
\end{array}
\right)
\ee

The natural question one may ask now is what is the meaning of the
reflection matrices of more general form. In particular, one can consider the
reflection matrix to be of the form (\ref{K})-(\ref{ABCD}) with arbitrary
$Q_0$ and $Q_2$. This results in removing the singularity $\lambda^4$ in the
last term of (\ref{scD}), i.e. there will be a general quadratic polynomial
of $\lambda^2$ instead. In brane terms, this looks like the four coinciding
D6 branes that formed the orientifold become split.
Therefore, this phenomenon could be naturally associated with the orientifold
splitting \cite{split}. The curcial difference with the cited paper,
however, is that there the splitting has emerged as a result of
non-perturbative corrections to the perturbative answer, while, in our
case, all the corrections are already taken into account.
Therefore, the amount of the splitting would provide another
non-perturbative parameter that can be related to
the Casimirs of the reflection algebra.

At the same time, the general reflection matrix of the form (\ref{Klinear})
(in the proper normalization)
\be\label{Klinear'}
K(\lambda)=\left(
\begin{array}{cc}
\alpha+{\delta\over \lambda}&\beta\\
\gamma &-\alpha+{\delta\over\lambda}
\end{array}
\right)
\ee
can be associated with a deformation of the orientifold with charge -4.
This deformation also resolves the singularity $\lambda^{-4}$ and, in a way,
"splits" the orientifold.

Moreover, it makes sense to consider even more general form of the reflection
matrix\footnote{The reflection matrix
(\ref{genK}) is of the most general form for a certain class of reflection
matrices \cite{genK}.}
that includes both (\ref{K})-(\ref{ABCD}) and
(\ref{Klinear'})\footnote{Note that $\alpha$ and $\delta$ in this reflection
matrix are central elements and, therefore, do not effect the number
of gauge moduli.}
\cite{Kuz2}
\be\label{genK}
K(\lambda)=\left(
\begin{array}{cc}
\alpha \lambda^2 +A_{1}\lambda +A_{0}+{\delta\over\lambda}&
\beta \lambda^2 +B_{0}\\
\gamma \lambda^{2} +C_{0}&D(\lambda)
\end{array}
\right),\ \ \ \
D(\lambda)=-A(-\lambda)
\ee
and ask for the corresponding brane picture. The simplest possible conjecture
would be that generalized reflection matrices correspond to combined
systems of different orientifolds. This point deserves further investigation.

\paragraph{{\large\bf Concluding comments.}}

We have shown that in order to describe the $N=2$ SQCD for different
classical groups within the integrability approach, one has to include into
the game non-trivial boundary reflection matrices corresponding to the
orientifolds in the brane approach. Parameters of the reflection matrices are
associated with positions of D4 and D6 in the brane set-up. We also
discussed generalizations of reflection matrices. The
deformation of the corresponding brane picture requires further
investigations. The simplest deformation is naturally associated with the
orientifold splitting therefore it would be very interesting to recognize
the reflection algebra behind the singularity structure of the
elliptically fibered Calabi-Yau threefolds.

It would be also interesting to construct the generalization of this
construction to $5d$ and $6d$ cases which looks quite immediate. Indeed, it
is known \cite{hd} that higher dimensional $SU(N_c)$ theories are described
by the XXZ and XYZ chains in, respectively, 5 and 6 dimensions. Therefore,
the only problem is to extend these results to the other classical groups,
i.e. to involve non-trivial boundary conditions. This means that one has to
construct the corresponding (trigonometric or elliptic) solutions to the
reflection equation\footnote{In fact, some solutions are already known,
see, e.g., \cite{RES}.}. Note that for constructing elliptic ($6d$) systems
of non-$A_n^{(1)}$ type one needs the standard numerical elliptic $r$-matrix
of $A_n$ type. This is of crucial importance since there are no
numerical elliptic $r$-matrices of non-$A_n$ type, which is an objection for
constructing the corresponding integrable systems in $N\times N$
representation.

The other quite immediate generalization of the proposed approach is to
consider the spin magnets given by Lax matrices of larger size $p$
(the so-called $sl(p)$ spin chains) and the corresponding reflection matrices.
The periodic $sl(p)$ spin chain has been considered in \cite{GGM1}. It is
shown to describe the $N=2$ SUSY YM theory with the gauge group being
the product of several $SU(N)$ groups and with bi-fundamental matter
hypermultiplets (see \cite{W}). In order to extend this to the other
classical groups, one merely needs to construct the corresponding solutions
to the reflection equation (\ref{RE}). The brane picture for this case is
discussed in \cite{orien,O6}.

Even more interesting problem would be to construct the generalization of the
scheme discussed in the paper to the exceptional groups $E_6$, $E_7$ and
$E_8$. In this case, we meet the new phenomenon of junction on the Dynkin
diagram. If we learn how to deal with this case, it would presumably
imply a possibility of associating an integrable spin chain with arbitrary
quivers, not obligatory of the Dynkin type. This would help to describe more
general SUSY theories that are a low-energy limit of string
theory compactified onto  manifolds  whose singularity structure is governed
by the corresponding quivers.

We are grateful to S.Gukov and S.Kharchev for discussions and to
S.Theisen for useful communications. The work is partially supported by
RFBR grants 98-01-00327 (A.Gor.), 98-01-00328 (A.Mir.),
INTAS grants 96-482 (A.Gor.), 97-0103 (A.Mir.),
the program for support of the scientific schools 96-15-96798 and
the Royal Society under a joint project (A.Mir.). A.Gor. thanks
Mittag-Leffler Institute where the part of this work was done for
the hospitality.

\end{document}